\documentclass{appolb}
\usepackage{epsfig}

\newcommand{\beq}{\begin{equation}} 
\newcommand{\eeq}{\end{equation}}   
\newcommand{\bea}{\begin{eqnarray}} 
\newcommand{\eea}{\end{eqnarray}}

\begin{document}
\date{\today}
\pagestyle{plain}
\newcount\eLiNe\eLiNe=\inputlineno\advance\eLiNe by -1
\title{New developments \\ in PHOKHARA Monte Carlo generator
\thanks{Presented by H.~Czy\.z 
at XXXI International Conference of Theoretical Physics Matter To The Deepest:
Recent Developments In Physics
of Fundamental Interactions, Ustro\'n , 5-11 September 2007, Poland.
 Work supported in part by 
 EC 6-th Framework Program under contract
  MRTN-CT-2006-035482 (FLAVIAnet),  TARI project RII3-CT-2004-506078
 and
  Polish State Committee for Scientific Research
  (KBN) under contract 1 P03B 003 28.}
}
\author{Henryk Czy\.z$^a$, Agnieszka Grzeli\'nska$^b$ and Agnieszka Wapienik$^a$
\address{ a: \ Institute of Physics, University of Silesia,
PL-40007 Katowice, Poland \\ b: \ Institut f\"ur Theoretische Teilchenphysik,
Universit\"at Karlsruhe,\\ D-76128 Karlsruhe, Germany
 }}
\maketitle

\begin{abstract}
 The present status of the physics program, which led to the development
 of the Monte Carlo event generator PHOKHARA is described. 
  The possibility of using the radiative return method 
 in various aspects of hadronic physics, 
 from the measurement of the hadronic cross section,
 to detailed investigations of  the hadronic dynamics is emphasized.
 New results are presented showing how to measure baryon form factors using
 the knowledge of their spin in baryon-antibaryon production with subsequent
 decay.
        
 \end{abstract}

\section{Introduction}
  The radiative return method proposed for the first time 
 to serve as 
 a tool in the hadronic cross section measurement \cite{Zerwas}
 requires Monte Carlo
 event generator(s) as theoretical
 input . The method relies on the factorization properties
 of the differential cross section with photon(s) emitted from the initial
 states (ISR) and the possibility to solve problems caused by
 the photons emitted from the final hadrons (FSR). 
The first tools to meet the expectations from experimental
 groups were developed (EVA \cite{Binner}, EVA4pi \cite{CK}) using the 
 structure functions method. That was a limitation, as not only the accuracy
 of the codes was not adequate to fulfill the experimental expectations,
 but also within that framework 
 more theoretical work is required to deal with possible
  double counting of the configurations, where the hard photon is emitted
  at low angles.
  The adopted approach in the continuation of the research
  program, mainly fixed order exact calculations, resulted in 
  the  development of the state-of-the-art Monte Carlo event generator 
  PHOKHARA\cite{Szopa,rest,Czyz:PH03,Nowak,Czyz:2004nq,PHOKHARA:mu},
 successfully used by the experimental groups  BaBar, BELLE and KLOE 
 working at meson
  factories.
 The hadronic cross section extracted using the radiative return method from
 the cross section of the reaction 
 with emitted
 photons gives  complementary information
 to the scan method, traditionally used for such purposes. It is needed
 to calculate, via dispersion relations,
  the anomalous magnetic moment of the muon 
 (for recent reviews see \cite{Jegerlehner:2007xe}) and the
 running of the electromagnetic coupling 
(for recent review see \cite{Jegerlehner:2006ju}).
 Profiting from the huge luminosities of the meson factories one can
 get valuable physical information, without building new accelerators and
  with accuracy competitive to traditional methods, as 
 demonstrated \cite{KLOE2} by the KLOE collaboration.
 As the method allows for the extraction of the hadronic cross section
 for energies from a production threshold to (almost) the nominal energy
 of the experiment, the B-factories have an access to data in the energy
 regions not covered previously by other experiments. Already now many
 new results exist, replacing the old measurements and/or covering  new
 energy regions, crucial for a precise evaluation of $(g-2)_\mu$ and
 $\alpha_{QED}(Q^2)$. The method originally developed for the hadronic
 cross section measurement has, however, broader applications and can be
 used to study the hadron dynamics. Work started in this direction
 in \cite{Nowak}, where it was shown how to extract nucleon form factors.
 It was continued in \cite{Czyz:2004nq}, where a method to test
 various models of the radiative $\phi$ decays was proposed.
 Working along this line, a newly
 published analysis \cite{Czyz:2007wi} shows how to use spin information
 on the decaying baryons to measure  phase differences of their 
 form factors.
 
 The Monte Carlo event generator PHOKHARA 6.0 was tested intensively at
 each stage of its development and the proved technical precision
 of the code is at the level of a small fraction (0.1-0.2) of a per mill.
 The precision of the theoretical formulae used in the code is currently
 about 0.5\% as far as the ISR is concerned. That accuracy is however
 not good enough to fully profit from more then 2 fb$^{-1}$ data collected
 by KLOE \cite {Ambrosino:2007nx,Leone:2006bm} at DAPHNE and further
 work is required to meet the growing expectations.

 The paper starts with a short description of the radiative return method in 
 Section~\ref{sec2}. The present status of the physics program
  for the precision hadronic
  physics with the PHOKHARA Monte Carlo generator is outlined 
 in Section~\ref{sec3}.  A
 short summary is presented in Section \ref{sec4}. 
%
\section{The radiative return method in short\label{sec2}}

The radiative return method relies on factorization properties
of the cross section with photon emitted from the initial electron
 or positron
  \begin{eqnarray}
 d\sigma(e^+e^- \to \mathrm{hadrons} + \gamma ({\rm ISR})) &=&
\nonumber \\ 
  &&\kern-50pt
 H(Q^2,\theta_\gamma) \  d\sigma(e^+e^-\to \mathrm{hadrons}, s=Q^2) \ ,
\label{factorization}
\end{eqnarray}
where $Q^2$ is the invariant mass of the hadrons.
 A similar factorization holds also, when more than one photon is emitted 
 from initial states.
 The function
 $H(Q^2,\theta_\gamma)$ (or a more complicated function for multi-photon emission),
 at relatively low energies of meson factories,
 is given with high accuracy by QED only and is thus well
 known. From this follows that a measurement of the
 differential (in $Q^2$) cross section of the reaction
 $e^+e^- \to \mathrm {hadrons} + \mathrm{photons}$
 allows for a cross
 section $\sigma(e^+e^- \to \mathrm{hadrons})$ extraction
 for energies from the 
 production threshold to almost the nominal energy of a given experiment.

\begin{figure}[ht]
\begin{center}
\epsfig{file=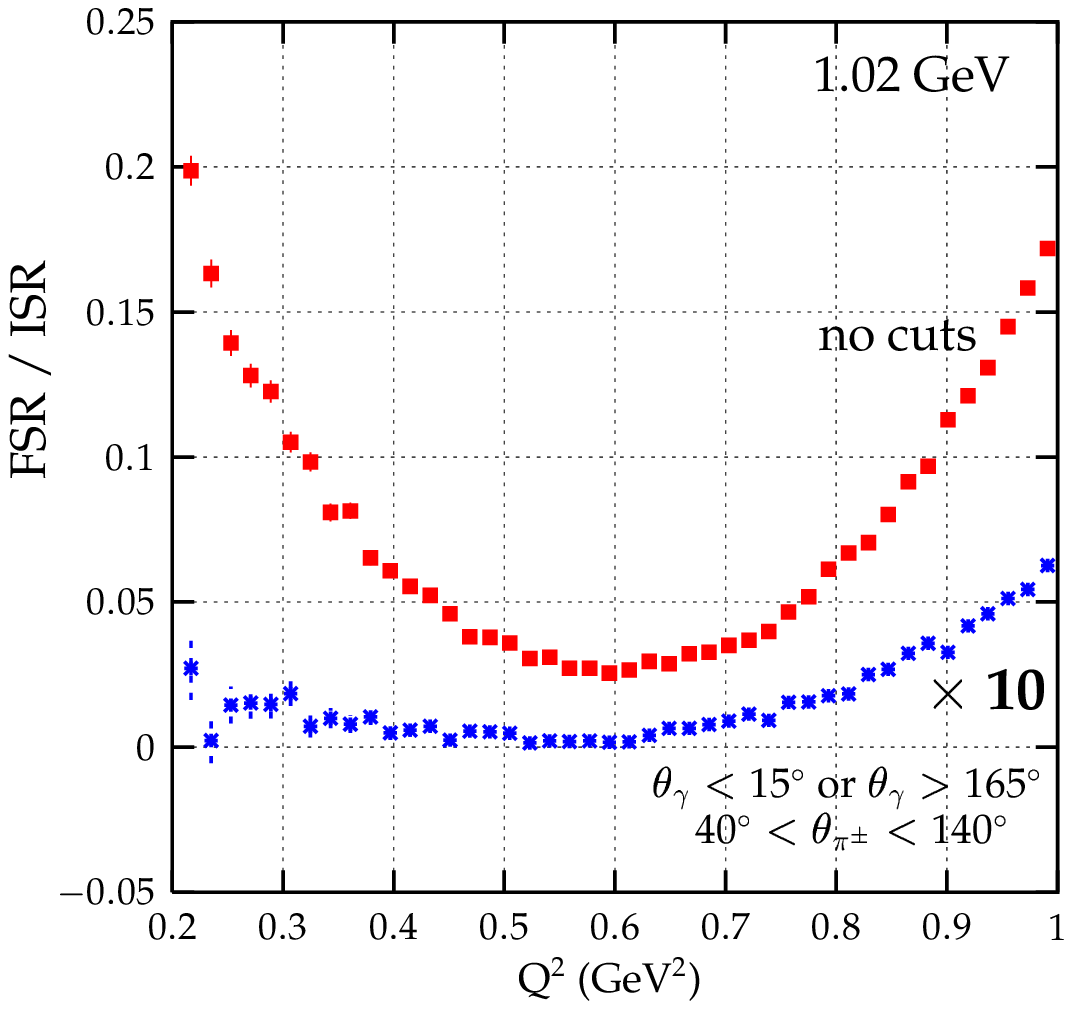,height=4.5cm,width=5.5cm} 
\epsfig{file=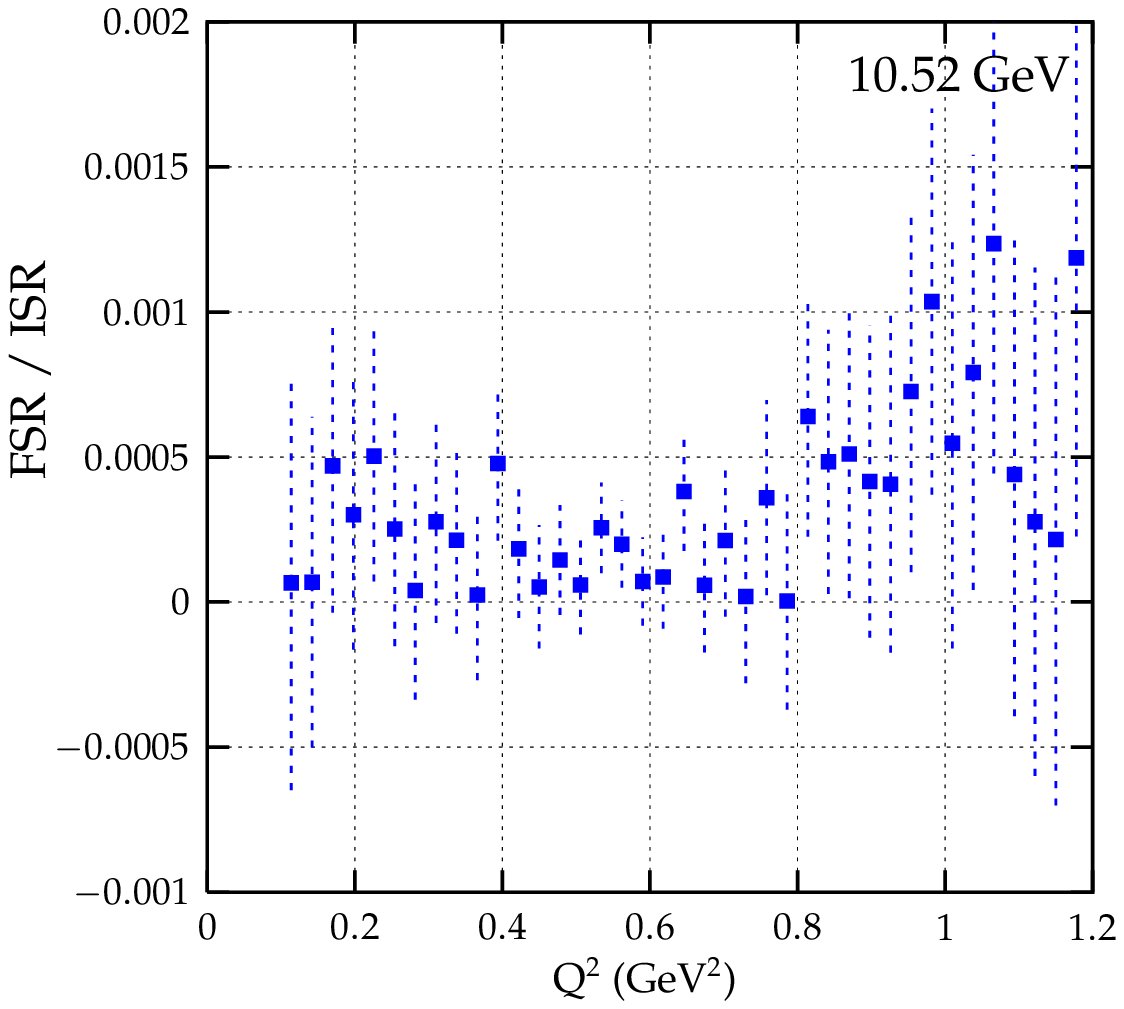,height=4.5cm,width=5.5cm} 
 \\
\caption{The relative leading order FSR contribution to the differential, in
 the invariant mass of the pion pair ($Q^2$), cross section of the reaction
  $e^+e^-\to \pi^+\pi^-\gamma$ for DAPHNE and B- factories energies. }
\label{fsrlo}
\end{center}
\end{figure}

%
%
 The presence of the contributions from photon(s) emitted from final
 hadrons has to be treated as a background. It has to be studied carefully
as the models of photon emission from hadrons are not well established.
The region of the hadron invariant masses, which is of main interest, is
 below 3 GeV and thus the role of the FSR at the $\phi$ factory DAPHNE is far more
 important then at B- factories (BELLE and BaBar). The reason is purely
  kinematical. To obtain a low invariant mass of the hadronic system
 one needs a very energetic photon emission at B- factories, which is not the case
 for the relatively low energy of $\phi$- factory. 
 The typical kinematical
 configuration at a B- factory is thus an energetic photon and hadrons going 
 in opposite directions.
 As the FSR contributions
 are enhanced only in the kinematical regions, where the 
 hadrons and photons directions overlap, the FSR is naturally suppressed
 at B-factories and it is not suppressed
 at  the $\phi$- factory DAPHNE, where special care
 is needed to deal with it. This is shown in Fig. \ref{fsrlo}, where the relative
 contribution of the leading order
 FSR corrections is plotted for the $e^+e^-\to \pi^+\pi^-\gamma$ differential
 cross section. Even if the FSR contribution at DAPHNE is sizable, when no 
 event selection is used, it is relatively easy to choose an event selection,
 which suppresses this unwanted background. A simple choice is also shown in 
  Fig. \ref{fsrlo}, but more sophisticated (and efficient) solutions can be
 found, as it was done by the KLOE collaboration 
 in their pion form factor measurement \cite{KLOE2}.
As the leading order contributions are suppressed at B- factories
 by the choice of
specific kinematical configurations, it is not a surprise that the next
 to leading corrections, which are not suppressed by that choice,
 are bigger then the leading corrections.  They amount up to a few
 percent and do depend on the invariant mass of the pion pair, and therefore have
 to be taken into account, when aiming for a precise measurement.  
 
%
%
 For DAPHNE energies the event selection used by KLOE to suppress the FSR contributions
 does not allow for a measurement of the pion form factor in the threshold region.
 That region might be important, as the contributions from low invariant masses are
 enhanced by the kernel function in the dispersion relations for the muon anomalous magnetic moment.
 Releasing the cuts allows for measuring in the threshold region, but in the same
 time the model dependence of the FSR contribution becomes a serious problem
 \cite{Achim_Ustron}.
 Detailed studies of the FSR, also for the  specific case of a $\phi$- factory,
 where the radiative $\phi$- decays play a role,
  were performed in many theoretical papers 
 \cite{Czyz:PH03,Czyz:2004nq,PHOKHARA:mu,Hoefer:2001mx} and that discussion
 will not be repeated here. One comment is, however, in order:
  the final solution to that problem can
 only come through close collaboration between theoretical
  and experimental groups
 and the analysis of both on peak and off peak data is crucial for the success
 of the program  \cite{Achim_Ustron}.
\section{An overview of the present status of the theory research program
 and new challenges\label{sec3}}
  The current version of the PHOKHARA 6.0 Monte Carlo event generator
 is a product of many years of theoretical investigations, calculations and code
 testing. It relies on virtual (+soft) radiative corrections to the ISR
  calculated in \cite{PHradcor} and ISR hard photon corrections calculated
 by means of the helicity amplitudes and implemented into an efficient
 Monte Carlo event generator in \cite{Szopa,rest}. The FSR at the 
next to leading
 order was investigated, and implemented into the event generator, for 
 muon and pion pair production in \cite{Czyz:PH03,PHOKHARA:mu}. In addition,
  very specific contributions to FSR, important
 at the $\phi$- factory DAPHNE and coming from radiative $\phi$- decays, was studied
 in \cite{Czyz:2004nq}. All the implemented parts of the developed code
 are tested to achieve a relative technical precision of few times $10^{-4}$.
 Independent tests were performed in \cite{Jadach:KKMC}, where the authors
 compared ISR contributions of muons present in PHOKHARA
 against the KKMC event generator \cite{Jadach:1999vf}. An excellent agreement
 between non exponentiated version of the KKMC and PHOKHARA was found, while
 higher order corrections, not implemented yet in the PHOKHARA, give at most
 two per mill contributions for the invariant masses of the muons relevant
 for the radiative return method.

 With PHOKHARA 6.0 one can now generate the following final states:
 $\pi^+\pi^-$, $\mu^+\mu^-$, $K^+K^-$, $\bar K^0 K^0$, $\bar p p$, $\bar n n$,
 $\pi^+\pi^-\pi^0$, $2\pi^+2\pi^-$,  $2\pi^0\pi^+\pi^-$, 
  $\bar \Lambda (\to \pi^+ \bar p) \Lambda (\to \pi^-  p)$, accompanied by
 one or two ISR photons. The FSR corrections
 are implemented only for  $\pi^+\pi^-$, $\mu^+\mu^-$ and $K^+K^-$, while
 in the prepared new release they are implemented also for 
   $\bar p p$ and $\pi^+\pi^-\pi^0$ final states. 
 The narrow resonance ($J/\psi$  and $\psi(2S)$) contributions to  
$\pi^+\pi^-$, $\mu^+\mu^-$, $K^+K^-$, $\bar K^0 K^0$ will also 
 be implemented there.
 The $\Lambda$ pair production and decays are implemented at the leading order
 only \cite{Czyz:2007wi}, but as the expected number of events is modest,
  the accuracy of the code
 is sufficient for the description of this process. The spin asymmetries
 and spin-spin correlations of the lambdas provide information about
 real and imaginary parts of the lambda form factors. To measure them,
 one needs only the information on the angular distributions of the 
 produced lambdas and pions coming from their decays \cite{Czyz:2007wi}.
 A step towards such a measurement was done by BaBar in \cite{:2007uf},
 but only a limited part of the information contained in the data was
 actually used.

  The model describing the four pion channels was improved, based
  on experimental information from BaBar \cite{Aubert:2005eg} and CMD2
 \cite{Akhmetshin:1999ty,Akhmetshin:2004dy} charged mode
  measurements  together with
  CLEO \cite{Edwards:1999fj}
  and ALEPH \cite{Schael:2005am} spectral functions.  
  The preliminary results indicate some isospin symmetry breaking effects,
  if one also adds information from the preliminary BaBar measurement
  of the neutral mode \cite{bb1}.

\noindent
  Even if relatively few final states are implemented in the distributed
 version of PHOKHARA, as compared to the plethora of available
 final states, the implementation of the missing channels by a potential
 user is not difficult, at least for the ISR part, due to the modular
 structure of the program. This was done for example by the BELLE collaboration
 and used in \cite{:2007ea}.

   The 0.5\% accuracy of the ISR corrections in PHOKHARA will soon be
  the biggest contribution to the error in the pion form factor extraction
  by KLOE. Inclusion of the leading logarithmic corrections from the second
  loop to one photon emission, leading logarithmic corrections at one loop level
  to two photon emission and implementation of the three hard photon emission
  is expected to bring the accuracy to the 0.2\% level and this will be the
   priority of the physics program of the group in the near future. 
  It will follow the accuracy improvement 
 \cite{Balossini:2006wc}
   of the BABAYAGA code used by KLOE for the luminosity measurement.  

%
%

%
\phantom{}
\vskip -0.5 cm
\section{Summary\label{sec4}}

The present status of the radiative return research program was outlined
and plans for the near future work towards further improvements of the
 PHOKHARA Monte Carlo generator were sketched. 
\vskip 0.5 cm


  The publication is based in a big part on results obtained in collaboration 
 with  
 J.~H.~K\"uhn, E.~Nowak-Kubat and G.~Rodrigo. The authors are grateful for 
many
 useful discussions concerning experimental aspects of the radiative
 return method to members of the KLOE, BaBar and BELLE collaborations, mainly 
 Cesare Bini, Achim Denig, Simon Eidelman, 
 Wolfgang  Kluge, Debora Leone, Stefan M\"uller,
 Federico Nguyen, Evgeni Solodov,  Graziano Venanzoni and Ping Wang.
%

\end{document}